\documentclass[twocolumn,amssymb,aps]{revtex4}
\usepackage{graphicx}
\usepackage[colorlinks]{hyperref}
\usepackage{amsbsy}

\begin{document}

\title{Spin-plasmons in topological insulator}

\author{D.\,K. Efimkin${}^1$}
\author{Yu.\,E. Lozovik${}^{1,2}$}\email{lozovik@isan.troitsk.ru}
\author{A.\,A. Sokolik${}^1$}

\affiliation{${}^1$Institute for Spectroscopy, Russian Academy of Sciences, Fizicheskaya 5, 142190, Troitsk, Moscow
Region, Russia\\
${}^2$Moscow Institute of Physics and Technology, Institutskii Per. 9, 141700, Dolgoprudny, Moscow Region,
Russia}%

\begin{abstract}
Collective plasmon excitations in a helical electron liquid on the surface of strong three-dimensional topological
insulator are considered. The properties and internal structure of these excitations are studied. Due to spin-momentum
locking in helical liquid on a surface of topological insulator, the collective excitations should manifest themselves
as coupled charge- and spin-density waves.
\end{abstract}

\maketitle

\section{Introduction}

In recent years, topological insulators with a non-trivial topological order, intrinsic to their band structure, were
predicted theoretically and observed experimentally (see \cite{Hasan} and references therein). Three-dimensional (3D)
realizations of ``strong'' topological insulators (such as $\mathrm{Bi}_2\mathrm{Se}_3$, $\mathrm{Bi}_2\mathrm{Te}_3$
and $\mathrm{Sb}_2\mathrm{Te}_3$) are insulating in the bulk, but have gapless topologically protected surface states
with a number of unusual properties \cite{Zhang}. These states obey two-dimensional Dirac equation for massless
particles, similar to that for electrons in graphene \cite{CastroNeto}, but related to real spin of electrons, instead
of sublattice pseudospin in graphene.

The consequence of that is a spin-momentum locking for electrons on the surface of strong 3D topological insulators,
i.e. spin of each electron is always directed in the surface plane and perpendicularly to its momentum
\cite{Hasan,Culcer}. The surface of topological insulator can be chemically doped, forming a charged ``helical
liquid''.

Collective excitation of electrons in such helical liquid were considered in \cite{Raghu}, where relationships between
charge and spin responses to electromagnetic field were derived. It was shown that charge-density wave in this system
is accompanied by spin-density wave. Application of spin-plasmons to create ``spin accumulator'' was proposed in
\cite{Appelbaum}. Also the surface plasmon-polaritons under conditions of magnetoelectric effect in 3D topological
insulator were considered \cite{Karch}.

In the present article we consider the properties and internal structure of spin-plasmons in a helical liquid. Within
the random-phase approximation, we derive plasmon wave function and calculate amplitudes of charge- and spin-density
waves in the plasmon state.

\section{Wave function of spin-plasmon}

Low-energy effective Hamiltonian of the surface states of $\mathrm{Bi}_2\mathrm{Se}_3$ in the representation of spin
states $\{|\uparrow\rangle, |\downarrow\rangle\}$ is $H_0=v_\mathrm{F}(p_x\sigma_y-p_y\sigma_x)$ for a surface in the
$xy$ plane, where the Fermi velocity $v_\mathrm{F}\approx6.2\times10^5\,\mbox{m/s}$ \cite{Zhang}. Its eigenfunctions
can be written as $e^{i\mathbf{p}\cdot\mathbf{r}}|f_{\mathbf{p}\gamma}\rangle/\sqrt{S}$, where $S$ is the system area
and $|f_{\mathbf{p}\gamma}\rangle=(e^{-i\varphi_{\mathbf{p}}/2},i\gamma e^{i\varphi_{\mathbf{p}}/2})^T/\sqrt2$ is the
spinor part of the eigenfunction, corresponding to electron with momentum $\mathbf{p}$ (its azimuthal angle in the $xy$
plane is $\varphi_{\mathbf{p}}$) from conduction ($\gamma=-1$) or valence ($\gamma=+1$) band. Many-body Hamiltonian of
electrons populating the surface of topological insulator is
$H=\sum_{\mathbf{p}\gamma}\xi_{\mathbf{p}\gamma}a_{\mathbf{p}\gamma}^+a_{\mathbf{p}\gamma}+
(1/2S)\sum_{\mathbf{q}}V_q\rho^+_{\mathbf{q}}\rho_{\mathbf{q}}$, where $a_{\mathbf{p}\gamma}$ is the destruction
operator for electron with momentum $\mathbf{p}$ from the band $\gamma$, $\xi_{\mathbf{p}\gamma}=\gamma
v_\mathrm{F}|\mathbf{p}|-\mu$ is its energy measured from the chemical potential $\mu$, $V_q=2\pi e^2/\varepsilon q$ is
the Coulomb interaction; $\rho^+_{\mathbf{q}}=\sum_{\mathbf{p}\gamma\gamma'}\langle
f_{\mathbf{p}+\mathbf{q},\gamma'}|f_{\mathbf{p}\gamma}\rangle a^+_{\mathbf{p}+\mathbf{q},\gamma'}a_{\mathbf{p}\gamma}$
is the charge density operator for helical liquid.

The creation operator for spin-plasmon with wave vector $\mathbf{q}$ can be presented in the form:
\begin{eqnarray}
Q_{\mathbf{q}}^+=\sum_{\vec{p}\gamma\gamma'}C_{\mathbf{p}\mathbf{q}}^{\gamma'\gamma}a^+_{\mathbf{p}+\mathbf{q},\gamma'}
a_{\mathbf{p}\gamma}.\label{Q}
\end{eqnarray}
This operator should obey the equation of motion $\left[H,Q_\mathbf{q}^+\right]=\Omega_qQ_\mathbf{q}^+$, where
$\Omega_q$ is the plasmon frequency. We can get solution of this equation in the random phase approximation at $T=0$
(similarly to \cite{Sawada}):
\begin{eqnarray}
C_{\mathbf{p}\mathbf{q}}^{\gamma'\gamma}=\frac{|n_{\mathbf{p}\gamma}-n_{\mathbf{p}+\mathbf{q},\gamma'}|\langle
f_{\mathbf{p}+\mathbf{q},\gamma'}|f_{\mathbf{p}\gamma}\rangle N_{\mathbf{q}}}
{\Omega_q+\xi_{\mathbf{p}\gamma}-\xi_{\mathbf{p}+\mathbf{q},\gamma'}+i\delta},\label{C}
\end{eqnarray}
where $n_{\mathbf{p}+}=\Theta(p_\mathrm{F}-|\mathbf{p}|)$ and $n_{\mathbf{p}-}=1$ are occupation numbers for
electron-doped helical liquid ($p_\mathrm{F}=\mu/v_\mathrm{F}$ is the Fermi momentum).

The plasmon frequency is determined in this approach from the equation $1-V_q\Pi(q,\Omega_q)=0$, where
\begin{eqnarray}
\Pi(q,\omega)=\frac1S\sum_{\mathbf{p}\gamma\gamma'}\frac{\left|\langle
f_{\mathbf{p}+\mathbf{q},\gamma'}|f_{\mathbf{p}\gamma}\rangle\right|^2
(n_{\mathbf{p}\gamma}-n_{\mathbf{p}+\mathbf{q},\gamma'})}
{\omega+\xi_{\mathbf{p}\gamma}-\xi_{\mathbf{p}+\mathbf{q},\gamma'}+i\delta}
\end{eqnarray}
is the polarization operator of the helical liquid, different from that for graphene \cite{CastroNeto} only by
degeneracy factor. The factor $N_{\mathbf{q}}$ in (\ref{C}) can be determined from the normalization condition
\begin{eqnarray}
\left\langle0\left|\left[Q_{\mathbf{q}},Q^+_{\mathbf{q}'}\right]\right|0\right\rangle=\delta_{\mathbf{q}\mathbf{q}'}
\sum_{\gamma\gamma'}D_{\gamma'\gamma}=\delta_{\mathbf{q}\mathbf{q}'},\nonumber\\
D_{\gamma'\gamma}=\sum_{\mathbf{p}}\left|C_{\mathbf{p}\mathbf{q}}^{\gamma'\gamma}\right|^2
(n_{\mathbf{p}\gamma}-n_{\mathbf{p}+\mathbf{q},\gamma'}),\label{norm}
\end{eqnarray}
($|0\rangle$ is the ground state), so that
$|N_{\mathbf{q}}|^{-2}=-S[\partial\Pi(q,\omega)/\partial\omega)]|_{\omega=\Omega_q}$. The quantities
$D_{\gamma'\gamma}$ in (\ref{norm}) can be considered as total weights of intraband ($D_{++}$) and interband
($D_{+-}+D_{-+}=1-D_{++}$) electron transitions, contributing to the plasmon wave function (\ref{Q}). Note that all
these formulas are also applicable to the case of graphene.

\begin{figure}
\begin{center}
\includegraphics[width=0.73\columnwidth]{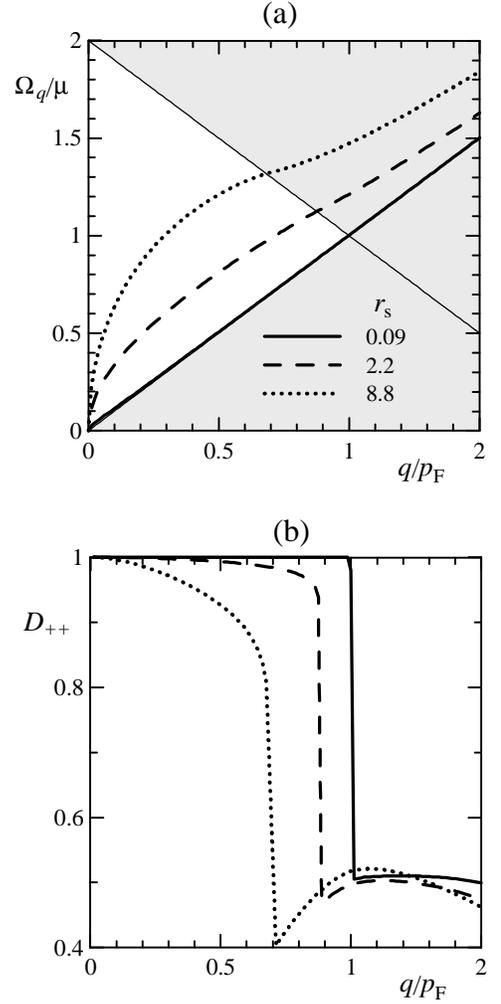}
\caption{\label{Fig1} Dispersions of spin-plasmon (a) and contributions $D_{++}$ of intraband transitions into its wave
function (b) at various $r_\mathrm{s}$. Continuums of intraband ($\omega<v_\mathrm{F}q$) and interband
($\omega+v_\mathrm{F}q>2\mu$) single-particle excitations are shaded in (a).}
\end{center}
\end{figure}

Spin-plasmon dispersion $\Omega_q$ and contribution of intraband transitions into its wave function are plotted in
Fig.~\ref{Fig1} at various $r_\mathrm{s}=e^2/\varepsilon v_\mathrm{F}$, where $\varepsilon$ is the dielectric
susceptibility of surrounding 3D medium. For $\mathrm{Bi}_2\mathrm{Se}_3$, $r_\mathrm{s}\approx0.09$ with
$\varepsilon\approx40$ for dielectric half-space \cite{Raghu} (for such small $r_\mathrm{s}$, the corresponding
dispersion curve approaches very closely to the upper bound $\omega=v_\mathrm{F}q$ of the intraband continuum). The
results for suspended graphene with rather large $r_\mathrm{s}=8.8$ (for $v_\mathrm{F}\approx10^6\,\mbox{m/s}$,
$\varepsilon=1$ and with the degeneracy factor 4 incorporated into $r_\mathrm{s}$) are also presented for comparison.
It is seen that the undamped spin-plasmon consists mainly of intraband transitions. When the dispersion curve enters
the interband continuum, the spin plasmon becomes damped and inter- and intraband transitions contribute almost equally
to its wave function.

\section{Charge- and spin-density waves}
The helical liquid in the state $|1_{\mathbf{q}}\rangle=Q^+_{\mathbf{q}}|0\rangle$ with one spin-plasmon of wave vector
$\mathbf{q}$ has a distribution of electron-hole excitations (\ref{C}), shifted towards $\mathbf{q}$. Due to the
spin-momentum locking, the system acquires a total nonzero spin polarization, perpendicular to $\mathbf{q}$. A similar
situation occurs in the current-carrying state of the helical liquid, which turns out to be spin-polarized
\cite{Culcer}.

\begin{figure}
\begin{center}
\includegraphics[width=0.7\columnwidth]{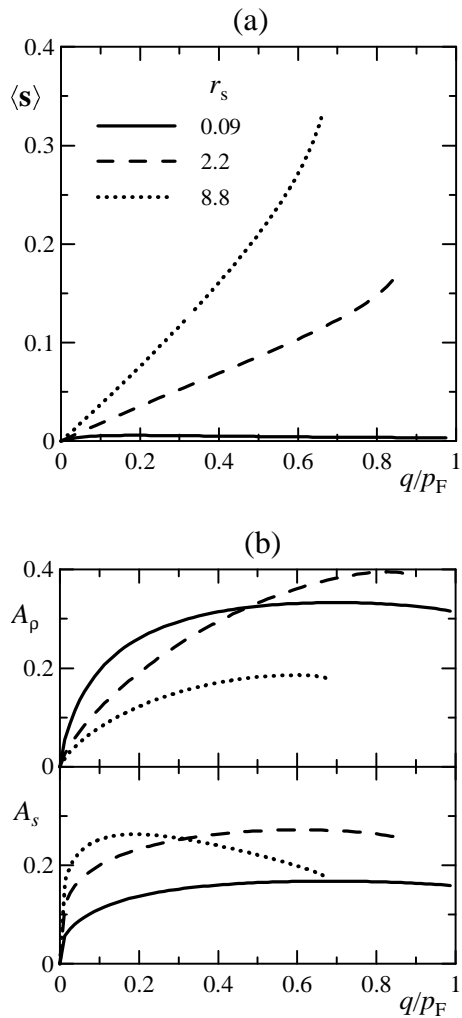}
\caption{\label{Fig2} Total spin polarization $\langle\mathbf{s}\rangle$ of the helical liquid in the one-plasmon state
(a) at various $r_\mathrm{s}$ and normalized amplitudes $A_\rho$ and $A_s$ of charge- and spin-density waves
respectively in the many-plasmon state (b).}
\end{center}
\end{figure}

Introducing one-particle spin operator as $\mathbf{s}=\boldsymbol\sigma/2$, we can calculate its average value in the
one-plasmon state $\langle\mathbf{s}\rangle=\langle1_{\mathbf{q}}|\mathbf{s}|1_{\mathbf{q}}\rangle$ as
\begin{eqnarray}
\langle\mathbf{s}\rangle=\sum_{\mathbf{p}\gamma\gamma'\tau}\left[\langle
f_{\mathbf{p}+\mathbf{q},\gamma'}|\mathbf{s}|f_{\mathbf{p}+\mathbf{q},\tau}\rangle
C^{\tau\gamma}_{\mathbf{p}\mathbf{q}}-C^{\gamma'\tau}_{\mathbf{p}\mathbf{q}}\langle
f_{\mathbf{p}\tau}|\mathbf{s}|f_{\mathbf{p}\gamma}\rangle\right]\nonumber\\
\times\left(C_{\mathbf{p}\mathbf{q}}^{\gamma'\gamma}\right)^*
(n_{\mathbf{p}\gamma}-n_{\mathbf{p}+\mathbf{q},\gamma'}).\label{st}
\end{eqnarray}
If $\mathbf{q}$ is parallel to $\mathbf{e}_x$, only the $y$-component of $\langle\mathbf{s}\rangle$ is nonzero. Its
dependence on $q$ at various $r_\mathrm{s}$ is plotted in Fig.~\ref{Fig2}(a).

Charge- and spin-density waves, accompanying spin-plasmon with the wave vector $\mathbf{q}$, can be characterized by
corresponding spatial harmonics of charge- and spin-density operators: $\rho^+_{\mathbf{q}}$ and
$\mathbf{s}^+_{\mathbf{q}}=\sum_{\mathbf{p}\gamma\gamma'}\langle
f_{\mathbf{p}+\mathbf{q},\gamma'}|\mathbf{s}|f_{\mathbf{p}\gamma}\rangle
a^+_{\mathbf{p}+\mathbf{q},\gamma'}a_{\mathbf{p}\gamma}$. Using, similarly to \cite{Brout}, the unitary transformation,
inverse with respect to (\ref{Q}), we can write:
$\rho^+_{\mathbf{q}}=SN_{\mathbf{q}}^*\Pi^*(q,\Omega_q)Q^+_{\mathbf{q}}+\tilde\rho^+_{\mathbf{q}}$ and
$\mathbf{s}^+_{\mathbf{q}}=SN_{\mathbf{q}}^*\boldsymbol\Pi^*_s(q,\Omega_q)Q^+_{\mathbf{q}}+
\tilde{\mathbf{s}}^+_{\mathbf{q}}$, where the operators $\tilde\rho^+_{\mathbf{q}}$ and
$\tilde{\mathbf{s}}^+_{\mathbf{q}}$ are the contributions of single-particle excitations and are dynamically
independent on plasmons. Here the crossed spin-density susceptibility of the helical liquid \cite{Raghu} has been
introduced:
\begin{eqnarray}
\boldsymbol\Pi_s(q,\omega)=\frac1S\sum_{\mathbf{p}\gamma\gamma'}
\frac{n_{\mathbf{p}\gamma}-n_{\mathbf{p}+\mathbf{q},\gamma'}}
{\omega+\xi_{\mathbf{p}\gamma}-\xi_{\mathbf{q}+\mathbf{q},\gamma'}+i\delta}\nonumber\\
\times\langle f_{\mathbf{p}+\mathbf{q},\gamma'}|f_{\mathbf{p}\gamma}\rangle\langle
f_{\mathbf{p}\gamma}|\mathbf{s}|f_{\mathbf{p}+\mathbf{q},\gamma'}\rangle.
\end{eqnarray}

The average values of $\rho^+_{\mathbf{q}}$ and $\mathbf{s}^+_{\mathbf{q}}$ in the $n_{\mathbf{q}}$-plasmon state
$|n_{\mathbf{q}}\rangle=[(Q_{\mathbf{q}}^+)^{n_\mathbf{q}}/(n_{\mathbf{q}}!)^{-1/2}]|0\rangle$ vanish, therefore we
consider their mean squares in $|n_{\mathbf{q}}\rangle$ after subtracting their background values in $|0\rangle$, i.e.
\begin{eqnarray}
\langle\rho_{\mathbf{q}}\rho^+_{\mathbf{q}}\rangle\equiv\langle
n_{\mathbf{q}}|\rho_{\mathbf{q}}\rho^+_{\mathbf{q}}|n_{\mathbf{q}}\rangle-\langle
0|\rho_{\mathbf{q}}\rho^+_{\mathbf{q}}|0\rangle\nonumber\\=n_{\mathbf{q}}S^2|N^*_{\mathbf{q}}\Pi(q,\Omega_q)|^2,\\
\langle s^\perp_{\mathbf{q}}(s^\perp_{\mathbf{q}})^+\rangle\equiv\langle
n_{\mathbf{q}}|s^\perp_{\mathbf{q}}(s^\perp_{\mathbf{q}})^+|n_{\mathbf{q}}\rangle-\langle
0|s^\perp_{\mathbf{q}}(s^\perp_{\mathbf{q}})^+|0\rangle\nonumber\\
=n_{\mathbf{q}}S^2|N^*_{\mathbf{q}}\Pi^\perp_s(q,\Omega_q)|^2
\end{eqnarray}
(only the in-plane transverse component $s^\perp$ of the spin $\mathbf{s}$ is nonzero in these averages). The
normalized amplitudes $A_\rho(q)=[\langle\rho_{\mathbf{q}}\rho^+_{\mathbf{q}}\rangle/n_{\mathbf{q}}S\rho]^{1/2}$ and
$A_s(q)=[\langle s^\perp_{\mathbf{q}}(s^\perp_{\mathbf{q}})^+\rangle/n_{\mathbf{q}}S\rho]^{1/2}$ of charge- and
spin-density waves are plotted in Fig.~\ref{Fig2}(b) ($\rho=p_\mathrm{F}^2/4\pi$ is the average electron density). The
``continuity equation'' for density and transverse spin, following from the spin-momentum locking \cite{Raghu},
requires that $\Omega_qA_\rho(q)=2v_\mathrm{F}qA_s(q)$, in agreement with our results.

\section{Conclusions}
We have considered microscopically spin-plasmons in helical liquid in the random phase approximation. The developed
quantum-mechanical formalism can be applied for a number of problems in spin-plasmon optics.

We calculated the average spin polarization, acquired by the helical liquid in a spin-plasmon state, as well as
mean-square amplitudes of charge- and spin-density waves, arising in this state. Coupling between these amplitudes,
caused by spin-momentum locking, was demonstrated. The interconnection between charge- and spin density waves can be
applied for constructing various spin-plasmonic and spintronic devices.

The work was supported by Russian Foundation for Basic Research, by Grant of the President of Russian Federation
MK-5288.2011.2 and by the Federal target-oriented program ``Research and scientific-pedagogical staff of innovational
Russia'' for 2009-2013. One of the authors (A.A.S.) acknowledges support from the Dynasty Foundation.

\end{document}